\documentclass{llncs}
\usepackage{llncsdoc}
\usepackage{pdfpages}
\usepackage{multirow}
\usepackage{pdflscape}
\usepackage{pifont}
\usepackage{changepage}
\usepackage{pbox}
\usepackage{booktabs}
\usepackage{multirow}
\usepackage{amssymb}
\usepackage{amsmath}
\usepackage{graphicx}
\usepackage{caption}
\usepackage{enumerate}
\usepackage{graphicx}
\newtheorem{mydef}{Definition}
\usepackage{flushend}

\begin{document}

\title{ A Role-Based Approach for Orchestrating Emergent Configurations in the \\ Internet of Things}

\author{Radu-Casian Mihailescu, Romina Spalazzese, Paul Davidsson, Clint Heyer}
\institute{Department of Computer Science, Malm\"o University \\
 Internet of Things and People Research Center, \\
 Malm\"o, Sweden\\
\{radu.c.mihailescu, romina.spalazzese, paul.davidsson, clint.heyer\}@mah.se}

\maketitle

\begin{abstract}

The Internet of Things (IoT) is envisioned as a global network of connected ”things” enabling ubiquitous machine-to-machine (M2M) communication. With estimations of billions of sensors and devices to be connected in the coming years, the IoT has been advocated as having a great potential to impact  the way we live, but also how we work. However, the connectivity aspect in itself only accounts for the underlying M2M infrastructure. In order to properly support engineering IoT systems and applications, it is key to orchestrate heterogeneous 'things' in a seamless, adaptive and dynamic manner, such that the system can exhibit a goal-directed behaviour and take appropriate actions. 
Yet, this form of interaction between things needs to take a user-centric approach and by no means elude the users' requirements. To this end, contextualisation is an important feature of the system, allowing it to infer user activities and prompt the user with relevant information and interactions even in the absence of intentional commands. 
In this work we propose a role-based model for \textit{emergent configurations of connected systems} as a means to model, manage, and reason about IoT systems including the user's interaction with them. We put a special focus on integrating the user perspective in order to guide the emergent configurations such that systems goals are aligned with the users' intentions. We discuss related scientific and technical challenges and provide several uses cases outlining the concept of emergent configurations.

\end{abstract}

\section{Introduction}
\label{intro}

Connecting everything that can benefit from being connected, from both digital and physical worlds, is becoming reality nowadays. IoT is already successfully adopted in several application domains such as environmental monitoring, healthcare service, transportation, inventory and production management among others \cite{li14,Atzori2010}. 
Recent studies show that IoT changed the way people use the Internet, mobile devices and sensors \cite{evans11}.

However, in order to reach its full potential, the IoT domain has to overcome a number of shortcomings.
On the one hand, in the enterprise space, the majority of IoT solutions suffer from the so-called 'cloud silo' problem. A common practice in today's IoT solutions is to have processes and data kept in separate data centers, making it difficult to access it or to interact with other systems. This design is often dictated by the companies'  business models or due to certain security concerns. Nonetheless, these silos can cause real issues by keeping data and services locked away and unavailable for an efficient use by external third parties. 
On the other hand, in the retail space of devising smart things as the building blocks of IoT, noticeably, \textit{things} are typically designed to follow a simple request-response pattern enacted by the user. This clearly imposes many restrictions in terms of the user experience and makes it impractical to specify requests and goals that concern more than one device at a time.

We aim to address these challenges encountered across the entire IoT spectrum in a unified manner.
The focus of our current work is to provide means to orchestrate heterogeneous \textit{things} in a seamless, adaptive and dynamic manner, such that the system can exhibit a goal-directed behaviour and take appropriate actions, according to the user's desires and intentions.  
As we shall see in the following sections, this raises many interesting research issues open for cross-pollination with the multiagent domain.
Informally, we term an \textit{Emergent Configuration (EC)} as a set of things with their functionalities and services that connect and cooperate temporarily to achieve a goal \cite{IDC_16_FR}.
Given the continuously changing character of many IoT systems, ECs can change unpredictably. This tendency is even more pronounced in domains that present pervasive characteristics such as mobile users, devices and sensors. Thus, it is important to provide the user with a coherent system at any point in time, capable to align to the user's context and goals. 

To this end, we put forward in this paper a general theoretical model for designing \textit{Emergent Configurations}. Our goal is to provide an encompassing framework where new mechanisms addressing the different aspects outlined by our model can be developed and deployed. Specifically, ECs set the stage for addressing and reasoning about three key categories of challenges that represent a barrier to effectively enable today's IoT systems:

\begin{enumerate}[(i)]
\item challenges in \textit{interoperability} - include aspects of how can devices operate across different types of networks, how to connect and deal with mobile devices, how to integrate devices of different hardware and software specifications, how to ensure ease of scalability?

\item challenges in \textit{coordination}: are concerned with how can devices collectively operate based on the user delegated goals to the system, how can devices separately adapt to runtime conditions while a consistent view of the system is maintained by the user, how to resolve data inconsistencies when retrieving the data from various sources?

\item challenges in \textit{user interaction} -  how to improve usability and provide novel ways of interaction with 
\textit{sets} of devices, how to provide a user-friendly interface for a variety of  devices, how to avoid overwhelming the user with data, how to improve usefulness and automatically identify relevant information, how to make the system proactive in supporting the user's tasks? 
\end{enumerate}

The reminder of the paper is organized as follows.
In Section \ref{sec:ecos} we present the basics of our proposed model characterizing \textit{Emergent Configurations}. In Section \ref{sec:usecases} we outline a representative EC use-case developed in co-production with industry partners. 
In Section \ref{sec:related} we discuss and point towards several key enabling technologies. 
Section \ref{sec:conclusions} concludes and draws future research directions.

\vspace{-2mm}
\section{Emergent Configurations for the Internet of Things}
\label{sec:ecos}  

\subsection{Basics of the proposed model}

\begin{figure*}[t]
  \centerline{
\includegraphics[scale=0.5]{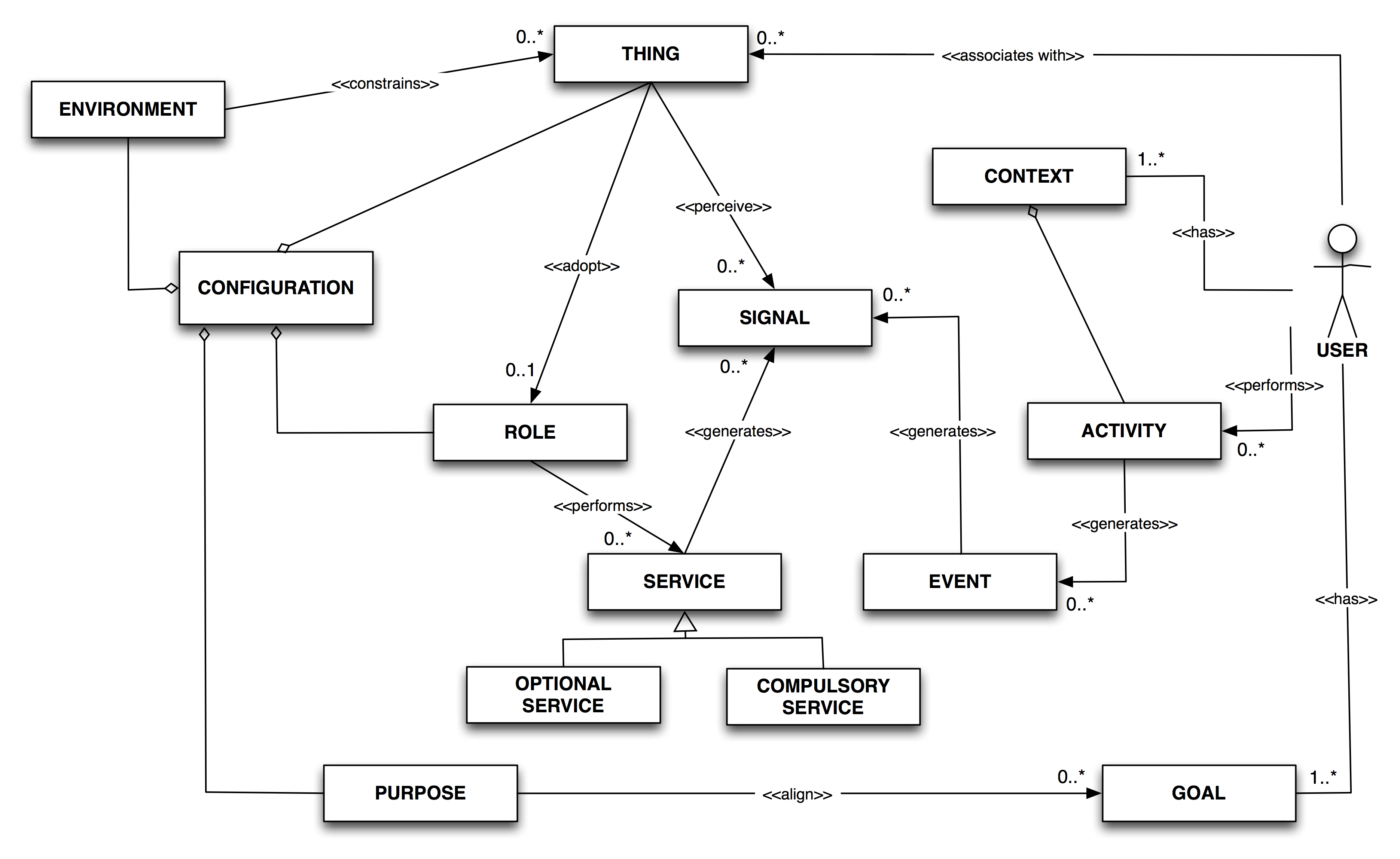}
  }
 \caption{Concept graph of the role-based model}
  \label{fig:overview}
 \end{figure*}

In this section we lay the groundwork for providing a model of emergent configurations in an IoT ecosystem. The proposed framework focuses in particular on the importance of \textit{roles} adopted by devices in various configuration settings. The aim is to provide a framework that is general enough to model different IoT scenarios and can capture the key aspects of adaptability to dynamic environments, while maintaining a coherent user perspective of the system according to its delegated objectives. This approach makes it easier to reason about integrating algorithms, which deal with different subproblems of the domain, as well as to reveal gaps in existing research.
In Figure \ref{fig:overview} we give an overview of the different elements considered in our model, which we describe hereafter.

\begin{mydef}
A \textit{connected thing} is an uniquely identifiable (embedded) computing system that has at least the capability to exchange data.
\end{mydef}

Basically, a \textit{connected thing} can represent anything, spanning from the simplest RFID tag or sensor, to actuators, devices, computing-enabled physical world objects, smartphones, computers, or even items of an increased complexity such as modern fully equipped self-driving cars.
It is further assumed that \textit{things} which are equipped with sensing capabilities can perceive, collect and index timestamped \textit{signals} received from the environment or communication messages from other \textit{things} in the environment. As a general approach, signal sequences registered over a certain time-frame are aggregated and classified to a higher-level representation, generating a new sequence of \textit{events}.
Similarly, in a hierarchical manner, the system can construct higher-level abstractions from the underlying sensing data.

\begin{mydef}
A linear sequence of signals or events $\epsilon_1, \epsilon_2, \cdots \epsilon_t$ obtained over the time period $t$ is perceived as (part of) the \textit{activity} performed by the \textit{user} in the environment. 
\end{mydef}

Suppose as an example a scenario where a smartphone is used to track signals from the accelerometer, gyro and GPS, inferring events such as accelerating speed, entering a park and continue at increased speed, in order to characterize a 'jogging in the park' activity.
The activity carried out by the user can be further enriched with additional information (e.g. time of day, weather, etc.), which captures the overall \textit{context} of the user \cite{Dey01,Cristea2013}. 

\begin{mydef}
 \textit{Context} is denoted here as the set of current statements about the user, which are either directly sensed from the environment or inferred. 
 \end{mydef}

In fact, there is an extensive body of work focusing on the area of activity recognition and context-awareness, as it pertains to the various sensing modalities (e.g. smartphones \cite{Kwapisz2011}, vision \cite{Poppe2010}, ambient sensors \cite{Mihailescu16,Mihailescu15}) and the different application domains (e.g. health \cite{Chernbumroong2013}, smart living spaces \cite{Mihailescu17}, sports \cite{Bulling2014}). 
For a comprehensive survey on context-aware computing we further refer the reader to \cite{Context_IoT,SURVEY_BETTINI_10}.
The underlying goal is to devise techniques that seamlessly acquire context information without the explicit intervention or feedback of the user. The use of semantics is one potential avenue for introducing meta-mechanisms that capture the relation between contexts and trigger the appropriate techniques for context inference, in a more general setting that can cover a wider spectrum of senors and domains. In addition, an important part of the context is the \textit{user profiling}, which retains the user's preferences. 

\begin{mydef}
The \textit{environment} $\mathcal{E}=\langle \Lambda,  \mathcal{T}\rangle$ specifies a set of constraints $\Lambda=\{c_1, \cdots, c_n\}$ which induce the set of things $\mathcal{T}$, providing a clear delimitation between the \textit{things} in $\mathcal{T}$ and those outside this set. Constraints are further differentiated into \textit{physical constraints} (e.g. devices at a certain location, communication ports, resource configuration) and \textit{virtual constraints} (e.g. communication protocols, restrictions over  virtual platforms, latency requirements). 
\end{mydef}

\begin{mydef}
A \textit{role} is a tuple $\Pi =  \langle \varphi, \chi,  \mathcal{S} \rangle$, which specifies an invocation function $\varphi : \chi \rightarrow \mathcal{S}$, which activates, for a given condition in the set $\chi$, an item $s \in \mathcal{S}$ from the restricted set of possible actions or service type interactions $\mathcal{S}$ available for this role. 
\end{mydef}

Items in the condition set $\chi$ can denote anything from incoming messages to perceived events or activities. Note that, a role can also be interpreted as a certain skill set $\mathcal{S}$, based on the actions and services that the role can perform.

\begin{mydef}
A \textit{configuration} is a tuple $\mathcal{C} = \langle \mathcal{R}, \mathcal{T}, \Delta, \beta, \mathcal{E} \rangle$, where $\mathcal{R}$ is a collection of \textit{roles},  $\mathcal{T}$ is a collection of \textit{things} and $\Delta : \mathcal{R} \rightarrow \mathcal{T}$ represents a mapping of roles to things. With $\beta$ we denote the \textit{purpose} of the configuration and $\mathcal{E}$ represents the environment. 
\end{mydef}

It is important to highlight that function $\varphi$ also specifies the \textit{type} of action or service invoked by a certain role $r$. On the one hand, we distinguishing between services $E(r)$ that are available to that respective role in the configuration, and services provided by this role, $P(r)$, to be accessed by other roles in the configuration. On the other hand, we differentiate between services $M(r)$ that are compulsory in the configuration and those that are optional $O(r)$. The set of compulsory services in a configuration $\mathcal{C}$ induces the set of compulsory roles $\Gamma_{\mathcal{C}}(\mathcal{R})$, required in order to run the configuration. Optional roles $\gamma_{\mathcal{C}}(\mathcal{R})$ can, for instance, be used to map to \textit{things} in the environment that are only temporary part of the configuration.

In other words, \textit{roles} provide the interaction constraints between the different entities in the configuration and can be assimilated in a broader acceptation to a workflow or (negotiation) protocol, that \textit{connected things} adopting different roles need to abide to.
A simple example would be having a \textit{configuration} with the purpose of determining the device with the highest QoS for a certain service among a list of possible candidates. This would entail prescribing the roles of service provider and service consumer, whereas the interaction could specify a simple  Contract Net protocol \cite{cnet}, which follows, on the provider side, a predefined sequence of broadcasting a call for bids, followed by receiving and evaluating bids and finally, determining the winner.

Going back to the overview in Fig. \ref{fig:overview}, the inferred activity of the user captures her underlying goals in a given context, which in turn dictates what roles her \textit{associated connected things} can adopt in a configuration. We use the term associated connected things in relation to the user to denote any means which interfaces the user with the configuration. They can range from something as trivial as a PIR sensor detecting presence of a person, to more advanced means of interaction, such as a smartphone, which can require some form of user feedback. In the situation when different parts of the configuration retain different information about the user, it is important to maintain a coherent view and to resolve potential data inconsistencies, when retrieving the data from various sources. Trust-based data fusion techniques \cite{Yan14} can represent a viable manner for prioritizing, processing and harmonizing data from heterogeneous sources. Note, that the purpose of the configuration needs to be aligned with the user's goals, otherwise its significance to the user would be meaningless. This brings about a number of interesting research questions, that situates the user as the focal point of decision making in the system. 

For instance, it is important to identify what is an appropriate way to specify and represent user goals. The use of ontologies could provide a way for representing, on the one hand, the user goals and on the other hand, the purpose of the configuration. If the user and the configuration's ontologies align, then we could reason that the configuration can provide a relevant function for the user. A dialog-based user-interaction can allow the user to guide the actual execution run of the configuration. Also, to what extent should the system use unobtrusive techniques in order to elicit the user's goals and obtain contextual information, and when should this information be explicitly specified based on the user's interaction with the system? There is an underlying trade-off here between the proactiveness of the system and the transparency of the process. If the system displays an increased level of autonomy, which is not clearly understood by the user, this can in fact hinder its usability and frustrate the user. In this regard, the recently introduced research avenue of UbiCARS \cite{mettouris2014}, which aims to characterize ubiquitous recommender systems, can bring interesting contributions to the concepts emphasized here. 

\subsection{Emergence}

Given that the interaction between \textit{connected things} in a configuration is mediated through \textit{roles}, another important aspect concerns the way according to which roles are established. It is important to allow the users or the system itself do modify the system behavior, in terms of prescribing new roles in a configuration or evolving more specialized roles to carry out certain tasks.

\begin{mydef}
Let $\mathcal{U}$ denote a user with goal $\mathcal{G}$. We term an \textit{Emergent Configuration}, the outcome of generating a configuration $\mathcal{C} = \langle \mathcal{R}, \mathcal{T}, \Delta, \beta, \mathcal{E} \rangle$, where the purpose $\beta$ of the configuration accommodates the goal $\mathcal{G}$ of user $\mathcal{U}$, given the environment $\mathcal{E}$.
\end{mydef}

In this sense, we make the distinction between \textit{weak} and \textit{strong } emergence. In the former case, we have \textit{static} roles and we are only concerned with identifying a suitable mapping $\Delta$ from \textit{roles} to \textit{things}. In effect, this represents an open-ended process where things can join the configuration on-the-fly, adopting certain roles, or may choose to leave the configuration at a certain point in time. This may also included situations where mobile software agents, which are performing certain roles, migrate from device to device in order to achieve the purpose of the configuration. For the later case, we assume that \textit{roles} are \textit{dynamic} and can change over time in order to better accommodate the goals of the user and the purpose of the configuration. A key feature of the system is its capability to adapt roles in order to correspond with the functionalities of the connected things adopting those roles. 

On the one hand, role adaptation can be triggered by explicitly receiving instructions from the user regarding the behavior of the configuration. Interaction design can play an essential part in this regard. On the other hand, the \textit{things} associated to the user, through continuous monitoring, can determine changes in the context or activities perform by the user, and therefore infer a new user goal that sets in motion at run-time a role adaptation procedure. In the more extreme case, we can envision a setting where either the configuration is built in a top-down manner, solely from specifying a certain \textit{user goal} (which also represents \textit{purpose} of the configuration) or alternatively, a collection of roles are evolved to serve a newly acquired purpose in a self-organizing fashion. This can be translated in a simple example by considering a smart home environment where the user is presented with a new type of functionality, such as begin able to use the smartphone in order to control an operate a smart TV. In this instance, the smartphone and the TV adopt the roles of master and slave respectively. 

\subsection{Managing emergent configurations}

Another aspect that has to be considered is linked to the way according to which a configuration is managed.
A configuration $\mathcal{C}$, where $\mathcal{R} =\Gamma_{\mathcal{C}}(\mathcal{R})$ and $\gamma_{\mathcal{C}}(\mathcal{R})=\emptyset$, meaning that configuration $\mathcal{C}$ includes only compulsory roles, is termed a \textit{centralized} configuration. For example, the \textit{leader} role can be compulsory for initiating a configuration and regulating the interaction between the other members of the configuration. Then, if all the roles specified by the configuration are compulsory, the configuration is a centralized one.
At the opposite end of the spectrum, in a \textit{decentralized} configuration, we have $\mathcal{R} =\gamma_{\mathcal{C}}(\mathcal{R})$ and $\Gamma_{\mathcal{C}}(\mathcal{R})=\emptyset$. Here, there is no device that assumes a compulsory role and the set of services provided by the configuration depends solely on the optional roles adopted by devices in the configuration at that time and the services they can receive and  be expected to deliver.
We term a \textit{hybrid} configuration one that involves both compulsory and optional roles ($\mathcal{R}= \Gamma_{\mathcal{C}}(\mathcal{R}) \cup \gamma_{\mathcal{C}}(\mathcal{R}), \Gamma_{\mathcal{C}}(\mathcal{R}) \neq \emptyset, \gamma_{\mathcal{C}}(\mathcal{R}) \neq \emptyset$), in the sense that compulsory roles are instrumental and all the other roles depend on the services provided by them.

The request for adopting a certain role can either be solicited by a member of the configuration (e.g. the leader advertises roles), or proposed by a device in the environment. This can take the form of a simple 'handshaking' procedure or, in more complex scenarios, it can imply that a negotiation occurs which can result in a service level agreement (SLA) \cite{sla}, that regulates future interactions.
A normative extension of roles using representations such as obligations and prohibitions could further be used to enforce certain device behaviors \cite{Criado2011}.
What is then an appropriate way to handle communication? An option inspired from MAS is to communicate using ACL (Agent Communication Language), a standard language for agent communication defined by FIPA (The Foundation of Intelligent Physical Agents) \cite{fipa}.
Also relating to implementation issues, we point towards model-driven engineering as a proposed support technology for ECs and the IoT, and regard  \cite{IDC_16_FR} as a possible architecture provided for self-adaptive systems.

\subsection{Discussion}
Finally, perhaps nearest to our work, we can relate the model proposed here to several efforts addressing the design of ontologies for ubiquitous and pervasive systems (e.g.  CoDAMoS ontology \cite{Preuveneers05}). However, the research in this area is focused on developing specialized ontologies dedicated to model specific domains, such as BOnSAI \cite{Stavropoulos12}, which addresses energy savings in smart building environments and has a particular emphasis on modeling devices, appliances and actuators. Another example is the OntoAMI model \cite{Santofimia2009}, which presents a simplistic semantic model for universal use across ambient intelligence applications, but fails to incorporate the \textit{user} into their approach.
Alternatively, in this work, we put forward a novel concept of \textit{Emergent Configurations} to denote a set of things with their functionalities and services that connect and cooperate temporarily to achieve a goal and provide a framework based on the notion of \textit{roles} to support reasoning and developing such systems. Moreover, we put special emphasis on the \textit{user} and explicitly account for the user's goals based on its current context and activity, in order to guide the emergence of configurations, such that the system's goals are aligned with the users’ intentions. 

It is important to point out a number of key benefits attained via our role-based approach. From a development standpoint, roles enable a decoupling between the interaction aspects, which prescribe how collaboration between devices occurs, and the internal computational processes of each device, which need not be exposed to other third parties. From the conceptual point of view, roles can be referred to as design patterns, in the sense that a set of roles that are used to fulfill the purpose of a configuration could further be (adapted and) reused for similar purposes. That is, roles can promote quite general, high-level solutions, which can then be parametrized and instantiated in various specific applications. Roles encapsulate certain functionalities, which equip devices with the necessary building blocks for establishing dynamic interactions in continuously changing environments and which can be directly related to the user's goal. In this manner, devices can dynamically adapt in a versatile manner to different coordination protocols and different functionalities by assuming various roles in different configurations.

\section{Use case of emergent configurations}
\label{sec:usecases} 

\begin{figure}[t]
  \centerline{
  \includegraphics[width=\textwidth]{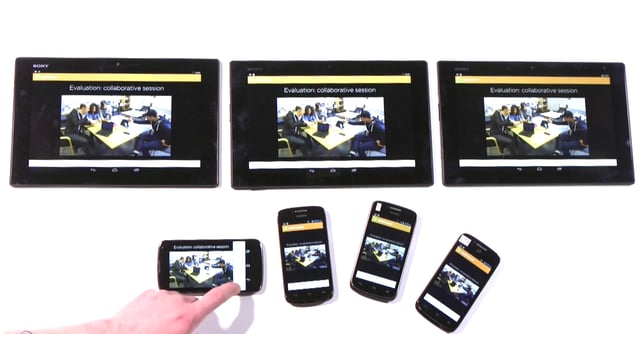}
  }
 \caption{Emergent Configuration use-case: Mesh Presenter}
  \label{fig:mesh}
 \end{figure}

In this section we introduce the "Mesh Presenter" prototype\footnote{https://vimeo.com/130228635} which has been developed as part of the ECOS project\footnote{http://iotap.mah.se/ecos/}, where we investigate how to model, interact with, and reason about ECs. 
We use this straightforward case study (see Fig. \ref{fig:mesh}) to emphasize the expressiveness of our formal model, which clarifies based on precise definitions how to dynamically integrate different entities and stakeholders to a configuration and how to regulate their behavior. 

The \textit{purpose} of the "Mesh Presenter" (MP) configuration is that of providing an \textit{on-the-fly} virtual collaborative workspace for sharing and generating content on smart devices (e.g. smartphone, tablets, projectors) co-located in a meeting room. In more detail, we describe the MP configuration according to the definitions introduced in Section \ref{sec:ecos}. Recall that roles represent one of the key underpinnings of our approach, providing a way to mediate interaction between different entities in the configuration in a dynamic manner.

Specifically, the MP configuration pertains that the set of roles $\mathcal{R}$ includes two types of roles. On the one hand, there is a compulsory role referred as $\Gamma(\mathcal{R})=\{presenter\}$ and an optional role termed $\gamma(\mathcal{R})=\{reviewer\}$. Each of these roles carries out a certain functionality, which can be described in terms of services. According to our model, we differentiate between two categories of services associated to each role. The set of services provided by the presenter include \textit{P(presenter) = \{share presentation; add/remove reviewer; enable presenter control\}}. This means that the reviewer role can request the presenter to execute any of these services. The presenter can grant or revoke access to any of the above-mentioned services to the reviewer. Given the collaborative context of this configuration, the reviewer can also actively participate in the content generation process by exposing the following service \textit{P(reviewer) = \{share content\}}. However, the presenter is the one who decides if the new data (e.g. image, slides) should be broadcasted on all devices within the configuration. For the sake of simplicity, in this usecase, having only two roles, the following sets coincide \textit{P(presenter) = E(reviewer)} and \textit{E(presenter) = P(reviewer)}, meaning that all services provided by one of the roles are accessible by the other. In addition, it is required that all services in \textit{P(presenter)} are implemented for this particular role, thus \textit{M(presenter)=P(presenter)} and $O(presenter)= \emptyset$. The same applies for the reviewer role.

In order to have a well-defined configuration several additional aspects need to be specified. We assume that users interact with the MP configuration by means of their associated smart devices. Importantly, the configuration environment $\mathcal{E}$ determines the set of devices that the presenter can allow to be part of the configuration. In this case, the inclusion constraint is based on a proximity rationale (i.e. presence in the meeting room). Also, the environment restricts the manner in which communication can take place, in this case being mesh connectivity. Contextualization also plays a significant part here. The real-time incoming data from the sensors of the device, such as a smartphone, are used to infer the user's activity. Once the \textit{context} of 'being present in a meeting' is recognized, this acts as a precondition for the reviewer role to invoke the service of joining the configuration. Similarly, when the presenter determines that another device has entered its environment $\mathcal{E}$ it will forward an inquiry to that device to join the configuration. User feedback is required in both of these instances in order to finalize the procedure. The preconditions for invoking all of the remaining services, such as sharing a presentation or granting presenter control to another device are all triggered based on a simple user interface, which basically summaries function $\varphi$, that is responsible for service invocation.

Note that, in the case of the MP configuration, we are dealing with a \textit{weak} type of emergence, as we are essentially assigning fixed roles to devices that are joining the configuration. Alternatively, a scenario that would exhibit a \textit{strong} emergence would entail that roles evolve and specialize over time, such that they can provide, new functionalities for the configuration, new services or new types of interaction (e.g. a new reviewer role that can connect remotely to the configuration). Finally, with respect to managing the configuration, we point out that MP is a \textit{hybri}d configuration as it consists of both compulsory and optional roles. A configuration can exist even in the absence of a reviewer, which can assume a temporary role, but not in the absence of the presenter which takes here a coordinator type of role.

\section{Enabling technologies}
\label{sec:related}

In our work, we are taking a \textit{user-centric} approach, by placing the users at the very center of the IoT ecosystem, which they have to manage in an efficient and effective way such that it can service their needs. We identify in the following, several key enabling technologies for IoT and point towards the contribution they can bring in the context of our model. Also, we relate them back to the three classes of problems outlined in Section \ref{intro}. Notably, we identify multiagent systems as arguably the main area of interest for advancing the EC concept presented in this work. Likewise, the user-centric approach advocated here, deploying a role-based solution for orchestrating IoT environments, provides an interesting perspective for new developments in MAS.

\subsection{Middleware}

The large majority of the research in IoT literature has focused thus far on dealing with the first class of challenges identified in the introduction, namely issues in M2M interoperability. In this sense, significant progress has already been made in terms of middleware platforms (see \cite{middleware_16,Delgado2013} for a comprehensive survey) designed to abstract hardware-specific issues and facilitate interoperability of various components, such as ad-hoc communication, discovery of new devices or establishing new communication links. These solutions also simplify the development process at the application level and exempts developers from handling implementation at the lower layers. In particular, service-oriented architectures (SOA) \cite{service,service_2} have emerged as a popular way for providing a middleware with standard protocols, common interfaces and well-defined components. This can be regarded as an approach by means of which connected things can easily interact and collaborate such that certain ECs are formed, according to the virtual constraints of the environment. Fog computing \cite{Bonomi2012} can also be regarded as an architectural blueprint for ECs, given its distributed infrastructure comprising of heterogeneous resources, which needs to be managed in a distributed fashion.

Another important aspect in our model that defines an \textit{emergent configuration} is the mapping function $\Delta$, which relates roles to things. A good candidate for implementing this function is given in \cite{Piette15}, where the authors propose a method for automatically deploying and configuring applications onto a heterogeneous hardware infrastructure, by taking into account the hardware properties and characteristics.
Alternatively, a multiagent system can act as an intermediary layer, providing the necessary level of autonomy for reasoning about adopting different roles in the configuration. At the same time, agent-based platforms, such as Calvin \cite{Per2015}, provide a lightweight distributed runtime, allowing the system to dynamically  deploy  agents according to requirements and hardware constraints.
Along the same lines, self-adaptive systems (SaS) \cite{SaS} can play an important role in enabling run-time reallocation, reconfiguration and  increasing reuse (e.g. thanks to model-driven technologies \cite{IDC_16_FR}).  
SaS are concerned with modifying the system's behavior at runtime in response to their perception of the environment or the system itself. In the last decade, the research in SaS has mainly focused on technical solutions, while less effort has been devoted to the human involvement that is the special perspective we advocate here. 

\subsection{Multiagent-based coordination}

Regarding the second class of challenges outlined in the introduction, in order to address the problem of device coordination, multi-agent systems (MAS) \cite{wooldridge09} is yet another important area we draw inspiration from in the realization of the concept of emergent configurations in IoT. In particular, we draw the parallel between ECs and an \textit{open} MAS, which assumes a high level of interaction between a large number of highly stochastic, heterogeneous and possibly self-interested agents operating within a dynamic environment. In such a system agents may form groups to solve more complex problems via cooperation, given that many tasks cannot be completed by a single agent because of limited resources or capabilities. A relevant example in this sense is given in \cite{Mihailescu13}, where the authors propose an online coalition formation scheme, having agents negotiate and formulate speculative coordination solutions, with respect to the estimated behavior of the system. This type of learning-based coordination can be instrumental for ECs in the way that roles evolve over time (representing in our scenario a case of strong emergence).

Importantly, agents running on devices can also act as a middleware in a configuration, enabling the interaction between devices, while at the same time adopting different roles in a versatile manner according to the user's goals. To tackle the problem of today's IoT platforms designed as vertical silos, we can envision a scenario, where at least one agent assumes a special \textit{broker} role, responsible for interfacing with each IoT platform, while the configuration emerges at a cross-platform level. This means that we could have for instance a situation where the configuration consumes data from a number of sensors registered to one IoT platform, then applies a data analysis component from another big data platform and visualizes the result on the projector that is managed by yet another local cloud platform. Notably, since a role-based approach is particularly aimed at providing a high-level description focusing on the interaction between \textit{things}, this has the advantage of simplifying and supporting the development of large scale applications.

In \cite{ossowski06} the authors introduce the notion of emergent coordination and discuss its potential for efficiently handling coordination in open environments. The authors advocate that a careful intentional design of a limited set of interaction rules and a common communication protocol could produce the desired emergent property, which is a higher level property caused by the interaction of its lower level components.
This brings about the question as to what are the key ingredients required in order to engineer a system such that it displays certain emergent properties, how does this relate to the types of emergence identified in ECs and what is the role of the user in this context?

Moreover, MAS allows to address challenges of autonomous and decentralized decision-making in a flexible manner, by decomposing complex tasks and assigning subproblems to loosely coupled components that interact and coordinate autonomously to solve system-level design goals. In a similar manner, ECs enhance open MAS by factoring in the user perspective. That is, ECs explore the extent of users involvement in the system not only as final user, but also as an active support for the system to guide or enable the execution of specific goals. An interesting preliminary work is \cite{garlan}, in which human participants are explicitly modeled and the authors reason about human involvement in self-adaptation, focusing on the role of human participants as effectors during the adaptation execution phase.

\subsection{Interaction Design}

Finally, from a user interaction perspective, again, people need to be involved not just as end-users, but as co-constructors of the system itself. A common strategy for co-construction has been designing for 'end-user programmability', such as rule based systems \cite{de2015homerules} and 'programming-by-example' \cite{dey2004cappella}. Here, devices are often used as the 'control panel' to represent, configure and control a complex system and its constituent components (\cite{ducheneaut2006orbital}, \cite{jin2014multi}). The disadvantage of this paradigm is that it draws attention away from the task-at-hand, making it difficult to continue the flow of activity \cite{roduner2007operating} and can suffer from device-centric rather than activity-centric design \cite{edwards2009experiences}. Thus, the challenge of designing IoT systems and applications that take a user-centric approach in order to guide the emergent configurations, such that the system's goals are aligned with the users' intentions, is still an open research question. 
Introducing roles takes one step further in this direction by means of binding the users' goals to certain roles assumed by their devices. Moreover, roles can be evolved in order to match changes in the user's goals. The framework described in this paper represents a vision that has to be tackled gradually, encompassing all of the identified challenges.

\section{Conclusions and future work}
\label{sec:conclusions} 

In this paper we defined and characterized emergent configurations of connected systems, as an approach to model, manage, and reason about collaborative IoT systems consisting of a number of autonomous yet coordinated devices and the user's interaction with them. 
We propose herein a role-based approach to ECs that captures a key missing aspect in today's IoT systems, namely, adaptability to dynamic environments, along with maintaining a coherent user perspective of the systems aligned with the user's objectives. 
The exploitation of roles at run-time enables \textit{things} to dynamically adopt specific roles in order to satisfy the user's goal or cope with dynamic environments. Our aim is to utilize a common framework, which can support  a wide range of IoT applications, instead of targeting specific scenarios, each of which can exercise particular aspects of the framework. We discussed scientific and technical challenges involving several research disciplines at the same time and we provided concrete cases showing how ECs can be used and help the IoT in practice.

As future work we plan to continue the development of the proposed framework, with a special emphasis on interaction design and the ways in which we can deliver an enhanced user experience providing functionality even in the absence of intentional commands, as well as on designing techniques to ease the development and deployment of ECs in real case scenarios. Moreover, in this work we are strictly concerned with generating a configuration with respect to the goal of a single user. In order to accommodate multiple user, conflict resolution mechanism (such as negotiation procedures) may need to be included, in the event that the goals of different users cannot co-exist in a given configuration. Regarding the evaluation, we would also like to identify and develop more full-fledged usecases that fully exploit the expressiveness of our model (e.g. configurations that display \textit{strong} emergence). It would be interesting to see the level of reusability we can attain by casting roles to new usecases and the impact of using the concept of roles throughout the analysis, design and implementation phases. Besides the development of additional cases and validation against domain-specific settings, the framework introduced here is also well suited for the integration, validation and comparison of already existing algorithm and mechanism or future proposals. Among them, we mention dynamic discovery of \textit{things} and services, goal adaptation mechanisms, approaches for configuration monitoring, context acquisition and inference, or novel ways for user interaction.

\vspace{5mm}

\textbf{Acknowledgments.} This work is partially financed by the Knowledge Foundation (KKS) through the Internet of Things and People research profile (Malm\"{o} University, Sweden). 

\bibliographystyle{plain}
\bibliography{biblio}

\end{document}